\documentclass{ws-ijmpc}

\begin{document}

\markboth{Andrzej Gecow}
{Structural Tendencies}

\catchline{}{}{}{}{}
%%%%%%%%%%%%%%%%%%%%%%%%%%%%%%%%%%%%%%%%%%%%%%%%%%%%%%%%%%%%%%%%%%%%%

\title{STRUCTURAL TENDENCIES - EFFECTS OF ADAPTIVE EVOLUTION OF COMPLEX
(CHAOTIC) SYSTEMS}

\author{ANDRZEJ GECOW\footnote{Research financed by government grant.}}

\address{Institute of Paleobiology Polish Academy of Science, \\
00-818 Warszawa Twarda 51/55 Poland\\
gecow@twarda.pan.pl}

\maketitle

\begin{history}
\received{Day Month Year}
\revised{Day Month Year}
\end{history}

\begin{abstract}
We describe systems using Kauffman and similar networks. They are directed functioning networks consisting of finite number of nodes with finite number of discrete states evaluated in synchronous mode of discrete time.
In this paper we introduce the notion and phenomenon of `structural tendencies'. Along the way we expand Kauffman networks, which were a synonym of Boolean networks, to more than two signal variants and we find a phenomenon during network growth which we interpret as `complexity threshold'. 
For simulation we define a simplified algorithm which
allows us to omit the problem of periodic attractors.
We estimate that living and human designed systems are chaotic (in Kauffman sense) which can be named - complex. Such systems grow in adaptive evolution.
These two simple assumptions lead to certain statistical effects i.e.
structural tendencies observed in classic biology but still not
explained and not investigated on theoretical way. E.g. terminal
modifications or terminal predominance of additions where terminal
means: near system outputs. We introduce more than two equally probable
variants of signal, therefore our networks generally are not Boolean networks. They grow randomly by additions and removals of nodes imposed on
Darwinian elimination. Fitness is defined on external outputs of
system. During growth of the system we observe a phase transition to chaos
(threshold of complexity) in damage spreading. Above this threshold we
identify mechanisms of structural tendencies which we investigate in
simulation for a few different networks types, including scale-free
BA networks.


\end{abstract}

\ccode{PACS Nos.: 11.25.Hf, 123.1K}

\section{Introduction}

In this short and condensed article we are going to introduce structural tendencies
which are effects of random adaptive evolution of complex systems.
Terminal modification and terminal predominance of addition are the
main examples of these tendencies. The former one means that a random change
of a system near system outputs has a higher probability of being 
accepted by adaptive condition than a change far from system
outputs. The latter one means that near system outputs more additions of
new nodes to network are accepted than removals of nodes in this
place. These tendencies are an equivalent of Naef's `Terminal
Modifications'\cite{Naef} and  Weismann's `Terminal Additions'\cite{Weismann1904} in the
evolutionary biology. Now these classic regularities are forgotten
together with comparative embryology \cite{Wilkins2002} only due
to lack of their explanation \cite{Gould1977}. Therefore our
investigations should have influence on biology \cite{bic}. Also
growth of the network is one of such structural tendencies and there is no need to assume it.

Complexity has as many descriptions and definitions
(e.g. \cite{Peliti88,Gell-Mann} or more recent \cite{Jost04,Jost06})   as different aspects and meanings. We define the complexity threshold during system growth as the phase transition to chaos. Over this threshold we observe in a simulation the
mechanisms of these structural tendencies. 
We use the term 'chaos' as Kauffman\cite{ooKauf} does. 
Systems exhibit chaotic behaviour when after a small disturbance of system the damage
statistically explodes and reaches a high level in equilibrium. Damage
is the difference between node output states in the disturbed and
undisturbed systems.

We estimate that the typical living or human-designed system grows
under adaptive condition and is chaotic. 
We treat the gene regulatory network as a strange exception. 

To describe such a system we use Kauffman's network
\cite{Kauf69,ooKauf} and similar ones but not in the autonomous case.
We do not use cellular automata ordered on lattice but a randomly
structured network. It is a directed network where each node is
influenced by $K$ other nodes or external inputs, and influences
$k$ other nodes or external outputs. We study $K=2$ or $K=3$.
The distribution of number $k$ depends on network type, $k$ may be fixed $k=K$ or
flexible in range starting from $k=0$. Each node influences others
by signals which are deterministic functions of the node's input signals.
The whole network has $m$ external input signals (in this paper
they are constant) and $m$ external output signals (in first step
$m=0$). System's fitness is defined as the number $b$ of output
signals which agree with an arbitrarily defined sequence of $m$ ideal
signals. In the network evolution changes are made randomly but
they form adaptive evolution only if they do not reduce system
fitness $b$.

Kauffman places living systems in phase transition between
order and chaos \cite{ooKauf} but he uses the Boolean network - the case of
Kauffman network with two variants of signal. He uses the assumption that the
two variants of signal have equal probability. We denote the   %p3
number of equally probable variants of signal as $s$. Note that using this 
parameter $s$ we know that these $s$ variants are equally probable and we will 
not repeat it every time. For more precision we assume in this all paper that
the `internal homogeneity'\cite{ooKauf} $P$ is the smallest. 
Only for $s=2$ the typical ($K=2$) big random system can exhibit ordered 
(opposite to chaotic) behaviour. 
If variants of signal are not equally probable (i.e. we cannot use description 
$s$), chaos can be avoided also for more than two variants. 

The mostly used case $s=2$  is also often not a realistic simplification. Typically real alternatives, especially when there are two alternatives, are not equally probable. This expected inequality is described using probability $p$ for one alternative \cite{Derrida86,Aldana03}.
We estimate that the typical occurrence of two unequally probable alternatives in descriptions of living and humane designed systems is an effect of our concentration on one of the possibilities which is special for us, and of collecting all remaining as second alternatives. Such a view needs another description than using $p$ for statistical mechanism because there are more than two real alternatives.

In the first step of our way we introduce $s>2$  and we show that
the parameter $s$ cannot be substituted by others (e.g. $K$ - see fig.~\ref{f1}). 
Typically $s=2$ is used for genome modelling (e.g. Ref.~\refcite{Kauf71,Serra07}). 
For genetic regulatory network where 1 is interpreted as active and 0 as inactive it seems adequate and gives results close to experimental data \cite{Wagner01,Serra04,Serrajtb07}.
However, when we are going to describe certain properties coded by genes or their mechanisms assessed using fitness we should remark that there are 4 nucleotide or 20 amino acids or other unclear spectra of alternatives, not only 2, therefore $s=20$ or $s=4$ seems much more adequate. Such strong simplification ($s=2$) can only be used if it does not significantly changed results, but we show, that it would do it. Stability or order in gene regulatory network results in properties assessed by natural selection in a long process where assumption of $s=2$ is probably not adequate.
Such a case described well by $s>2$ differs from one described by $p$ and $1-p$ 
in the statistical mechanism and its result. E.g. for extreme $p$ and small %p3
$K>2$ order is expected \cite{Derrida86,Aldana03} but not for $s>2$.
For $s>2$ a system should be chaotic which our coefficient $w=k(s-1)/s$
of damage propagation shows easily. However, to be chaotic a system
also needs to be big enough. In the second step our complexity threshold
shows how big it should be. This depends on the network type. Our
complexity threshold can be easily detected in the reality and in
the simulation.

In the last, third step we investigate structural tendencies in the
adaptive evolution of chaotic system (above the complexity threshold on
system size axis, with $s>2$). In our model we define fitness $b$, which is
needed in adaptive evolution, only on the system external outputs, not
on all the node states as in Kauffman model. Therefore our system in the third step
is not an autonomous one ($m>0$, in simulation we use
typically $m=64$). In the investigation of complexity threshold in
second step we also prefer such a form of the system to be
appropriate for the third step. Only in  the first step the considered
system is autonomous ($m=0$). In our fitness there are no local
extremes. To keep fitness not maximal but high and constant the
ideal vector of output signals is changed accordingly.

For simulation we use our simplified algorithm which allows us to
obtain one particular vector of output signals instead of a periodic
attractor. It gives correct answers for our statistical questions 
and practically allows to investigate the emerging structural tendencies.

\section{Coefficient $w$ of Damage Propagation}

% \& its basic parameters - number of signal variants $s$ and node degree $k$}

We suggest that an interpretation needs
usually $s>2$ which keeps models in the chaotic area. Kauffman started
from famous Ref.~\refcite{Kauf69}, used Boolean network and $s=2$. Using
such assumptions he concluded that the best place for systems to
adapt is the `phase transition between chaos and order' where the
`structural stability' occurs. 
Structural stability can be understood as the ability to small changes (see 'small change tendency' in the end of ch.5). This ability can be different in different areas of network. We differentiate between areas using distance to external outputs of network but Kauffman cannot do it using autonomous networks. He uses properties of the whole network, therefore he need order regime to obtain ability to small changes.
The phase transition to order when $s=2$ occurs \cite{Derrida86} between
$K>2$ and $K=2$.  The number $K$
of node inputs, which was typically  fixed \cite{fix} in considered
networks, is equal to the average  $k$  - number of node outputs.
We also use constant $K$, however, in more recent works \cite{Aldana03,Kauf04,Iguchi07} the in-degree distribution and out-degree distribution are considered.

\begin{figure}[ph]
\centerline{\psfig{file=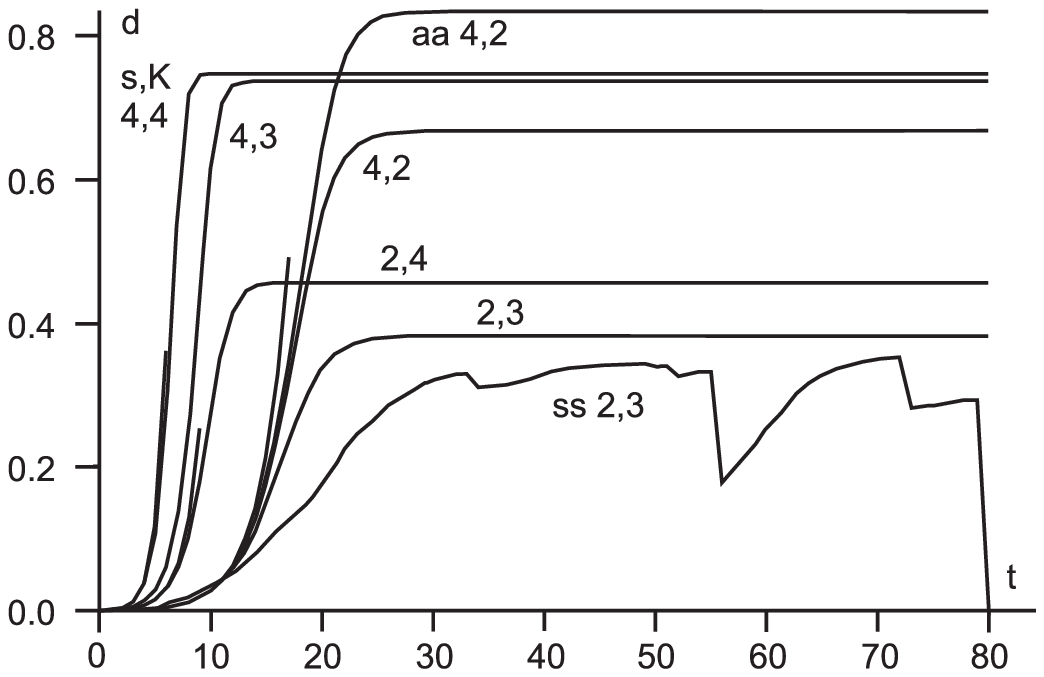,width=7cm}}
\vspace*{8pt}
%\begin{figure}
%\sidecaption[t]
%\includegraphics[height=4.5cm]{fig1.eps}
\caption{Theoretical damage size for Kauffman networks for
different $s$ and $K$ denoted as $s,K$ and for $aa4,2$. 
Approximation as $d=w^t$ of beginning part of damage size is shown for 4,4 
and 2,4 separately and together for $aa4,2$; 4,2; 2,3 which give 
the same $w=1.5$.   %p4 fig.1
Note the big difference of damage equilibrium for different $s$ but small 
for different $K$. 
Simulation result for $ss2,3$ is very similar to $sf2,3$ and $sf4,2$, where 
different speed is visible.}
\label{f1}      
\end{figure}

The coefficient of damage propagation $w=k(s-1)/s$  shows how many
output signals of a node are changed on average if its input state
is changed. It needs randomly defined deterministic functions of
nodes. Damage $d$ is a part of nodes with changed output state.
System is big if it consists of large number $N$ of elements. When
the damage avalanche is still small and the range of interactions
spans a whole and big system then probability of more than one
changed input signal is also small and damage $d$ is well
approximated by $w$ as $d(t)=d_0w^t$ (fig.~\ref{f1}). In this critical
time period $w$ describes the damage multiplication on one node.
If $w>1$ then the damage should statistically grow and spread on a
large part of  a system. It is similar to the coefficient of
neutron multiplication in a nuclear chain reaction - we have less
than one in a nuclear power station, for values greater than one an
atomic bomb explodes.

Note, that $w=1$ appears only if $k=2$ and $s=2$. Both these
parameters appear here in their smallest, extreme values. The
case $k<2$ is sensible for particular node but not as an average in a
whole, typical, randomly built network, however we can find the case $K=1$ in Ref.~\refcite{ooKauf,Wagner01}. For all other cases where $s>2$ or $k>2$ we have $w>1$.

The number $s$ of equally probable variants of signals is the next
main parameter of a system, like Kauffman's $K$ - number of
element's inputs and $P$ - the `internal homogeneity' in Boolean
functions. These parameters define a system as chaotic or ordered.
Note that parameters $s$ and $P$ work in opposite direction when they differ from their typical value - the smallest one. Higher $s$ causes chaos but higher $P$ allows to avoid chaos, however both of them are connected to the problem of equal probability of two variants of signal. They describe different aspects of this idealisation. 
The simple and intuitive coefficient $w$ may substitute two of
them ($s$ and $K$) in this role but this is only the first approximation. 
We have shown in fig.~\ref{f1} different levels of damage equilibrium for different $s$ but the same $w$. 
As we will show later, the type of network, i.e. distribution of node degree $k$, 
and number $N$ of node in the system, 
also has an influence on a place of system on the chaos-order axis , 
and this influence depends on $s$ and $k$ in other way than $w$.

When the damage is still small the probability of its fadeout is not
to be neglected. Later, it practically cannot fade out, cases of
more than one changed input signal happened more often and the
multiplication factor of damage decreases to one. Damage reaches a stable
level. See fig.~\ref{f1}, which we have calculated in the way described in
Ref.~\refcite{ooKauf}, expanded to the case $s>2$. E.g. for $K=2$ we
obtain $d_2=w(d_1-d_1^2/2)$   where, for small $d_1$, we can
neglect the second element.

In the Kauffman networks all $k$ outputs of a node transmit the same
signal - it is the state of the node, the value of a function. To
understand the coefficient of damage propagation, we must average
by the nodes. It is much simpler and more intuitive if each output
of a node has its own signal to transmit. In such a case, the function
value is a $k$-dimensional vector of signals. Due to function
uniformity it is useful to fix $K=k$. I have introduced such a
network in Ref.~\refcite{paris,hof,krab} where I have named it `aggregate
of automata'.

\section{Differences in Damage Behaviour for Different Network}
% types} %byl to 3 rozdz - ok

\begin{figure}[ph]
\centerline{\psfig{file=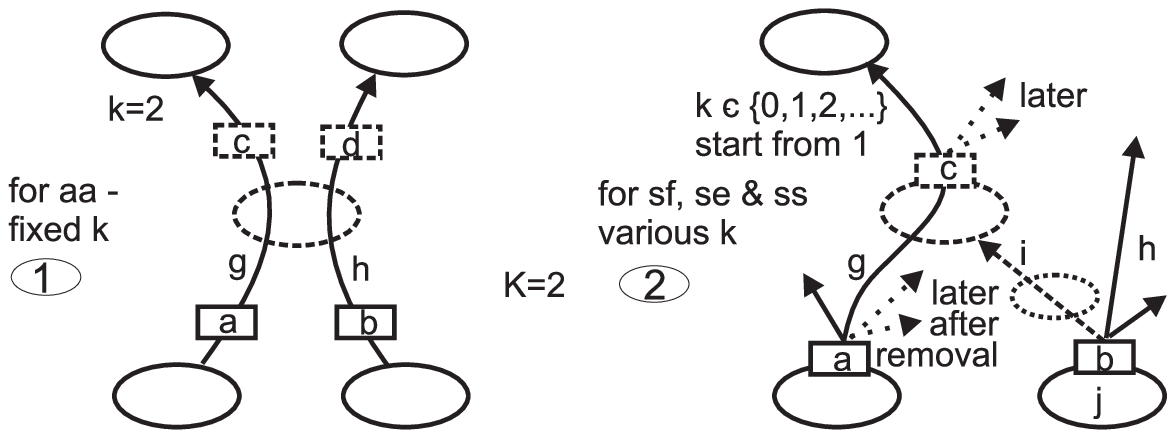,width=9cm}}
\vspace*{8pt}

%\begin{figure}
%\centering
%\includegraphics[height=3.2 cm]{fig2.eps}
\caption{Aggregate of automata (1) and Kauffman network (2)
changeability patterns for $K=2$. In  case of addition,  links $g$
and $h$ and function of node are drawn. Node $j$ is drawn instead of
link $h$ for $ss$. Removal only needs a draw of a node to remove.
For node removing in $ss$ and $sf$ the outgoing links, which were
added after addition of this node to the network, are moved to the
start node of link $g$. Node added on link $i$ remains a $k=0$ node while
removing, therefore for adaptive growth with removing the $sf$
network pattern for addition is supplemented by drawing  node $j$
additionally ($se$ network). For $K>2$ additional inputs are
constructed like the right ones. The $ak$ network is maintained as
$aa$ but there is only $c$ output signal.}
\label{f2}      
\end{figure}

However, Kauffman's formula gives an useful ability to differ $k$ and to
investigate networks types which differ in distribution of node
degree $P(k)$. Expectations shown in fig.~\ref{f1} are independent of the
type of a Kauffman network. We check them in simulation for a few
types of networks. We denote the network type by two letters: `$er$' -
old, typical, `random' Erd\H{o}s-R\'enyi \cite{er,Kauf69,ooKauf,Iguchi07}; 
`$sf$' - Barab\'asi-Albert scale-free network \cite{sf99,sf03}; 
`$ss$' - single-scale \cite{ss}; `$aa$' - aggregate of automata
\cite{paris,hof,krab}; `$ak$' - like $aa$, but $ak$ uses Kauffman
function formula (one output signal for all k outputs). The `$er$' network does not grow 
- it is built for fixed $N$. The `$aa$' and `$ak$' in order to add a new node need to
draw $K$ links, which are broken and their beginning parts become
inputs to the new node and the ending parts become its outputs
(fig.~\ref{f2}.1). For `$ss$' new node connects with the node present in
the network with equal probability for each one.  For `$sf$' new
node connects with another one with a probability proportional to its $k$  - 
node degree
(fig.~\ref{f2}.2).  We first draw one link for $sf$ and $ss$ and then we break
them like for $aa$ and $ak$ to define one output and the first
input. For $sf$ type at least one output is necessary to 
participate in further network growth. If $K=2$, then only one
input follows the rules but it is enough to obtain correct $P(k)$ ($k$-number 
of node outputs treated as node degree) distribution characteristic for these 
network types. 

Damage spreading in scale-free networks has often been investigated (e.g. 
Ref.~\refcite{Crucitti2004,Fortunato2005}) in nondirected networks.
Related studies based on complex computational networks have been 
conducted in Ref.~\refcite{nowostawski}. 
A directed scale-free network was used in Ref.~\refcite{Stauffer2006} %p6
preceded by Ref.~\refcite{Stauffer2004,Jaco2005}. These networks describe 
opinion agreement process and are not similar to Kauffman networks. 
Dynamics of Boolean networks 
with scale free topology were studied by Aldana \cite{Aldana03} and Kauffman 
\cite{Kauf04}, now Iguchi et al.  \cite{Iguchi07}. They look for difference 
between the dynamics of $er$ (here called: RBN) and the scale-free random 
Boolean network (SFRBN). Here $s$=2, flexible $k$ and $K$ are used, therefore 
their networks differ from our $sf$.

We use our own simple algorithm for statistical simulation of damage
in synchronous mode, where we only calculate nodes with changed
input state, not as in classic method two systems - changed $A$ 
and undisturbed $B$ \cite{Jan94}. We ignore the remaining input signals, 
they can be e.g. effects of feedback
loops. Each node output is calculated only once \cite{krab}.  We often
do not use a concrete function for node either - if the input state is
changed, then the output state is random. Such an algorithm works fast
and gives correct statistical effects.
Intuition behind this algorithm can be found when we imagine a network without feedbacks, where each node output  is equal to the function value of current node inputs. In such a case for calculating node with changed input signal we can use the old input signals as the remaining ones. After a finite number of steps the process will stop. The damaged part will become a tree, and all the node states will be equal to function value of current node inputs as was the case at the beginning. In the case with feedbacks sometimes an already calculated node gets a damaged input signal for a second time, but for measuring the statistical effect only it is not necessary to examine its initiation for the next time. 

When the network achieves $N=2000$ nodes we stop the growth and we
start to initiate  damage. As  damage initiation we change the output state in turn 
for each of the nodes. In the first few steps the damage can fade
out, but $w>1$, therefore on average the damage grows. Later the damage is
too large to fade out. During its growth there are less and less
nodes which are not reached by damage yet and damage slows down
then stops the growth. In our simplified algorithm it looks like
fadeout on nodes, which have been already affected by damage but
it corresponds with a stable equilibrium level, which appears at
the end of curves in fig.~\ref{f1}.

We have simulated the cases described as $s,K$: 2,3; 4,2; 4,3.
Each simulation consists of 600 000 damage initiations. Networks
are autonomous (without external inputs and outputs). When we
compare effects of simulation to the theoretical curve in fig.~\ref{f1} for
e.g. $sf$ and $ss$ 2,3 or $sf$ 4,2 (fig.~\ref{f1}. case $ss$ 2,3) we can
identify a few independent fractions of summarized processes which have 
different tempo of damage growth. 
In our simplified algorithm there are no data for time
steps later than `pseudo fadeout', therefore we can observe 
a slower group of processes. For $sf$ and $ss$ this tempo is strongly 
connected with the time when the damage reaches the hubs.  
For $er$ and obviously $ak$ and $aa$ there are no hubs and obtained curves 
are much more similar to the theoretical one shown in fig.~\ref{f1}.

In the distribution of damage fadeout in time dependency there are
two peaks: one for real fade out in the first steps and the second one
for `pseudo-fade out'. For networks with wide range of $k$ like
$sf$ and $ss$ with great fraction of $k=1$ (near $\frac{1}{2}$)
the probability of early fade out is much greater, especially for
small $s=2$. Here hubs are present, they decrease the average $k$ for
the remaining nodes which helps the damage to fade out before the first
hub is achieved. At the opposite end (only of Kauffman mode)  lies $ak$
4,3 where $k=1$ and hubs are absent and $w$ is high and equal
for all nodes. In such a case, the early fadeout is very small and
most of the damage grows up to the equilibrium level.

Different tempo of damage spreading causes the wideness of these
peaks and lack of sharp boundary between them. These peaks are
much  narrower and separated by a long period of exact zero
frequency if distribution of damage fadeout is shown in damage
size variable. In such a description the second peak shows exactly 
the equilibrium level and only the case $sf$ 2,3 appears extreme.

\begin{figure}[ph]
\centerline{\psfig{file=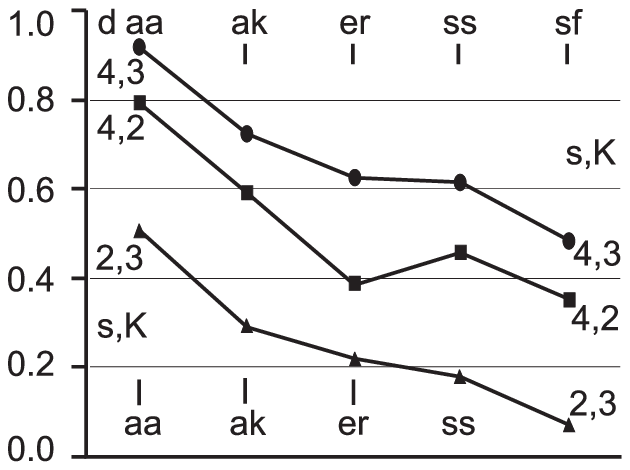,width=5cm}}
\vspace*{8pt}
%\begin{figure}
%\sidecaption[t]
%\includegraphics[height=3.6 cm]{fig3.eps}
\caption{Average damage size for different parameters $s$ and $K$:
4,3 4,2 and 2,3 and different network types: $aa$, $ak$, $er$, $ss$ and $sf$. 
The $er$ network is the only one with $k=0$. 
Note, that for 4,2 and 2,3 the coefficient $w=1.5$. 
We cannot delimit ourselves to only one of  the parameters $K$, $s$, $w$
or one network type. The shown points have 3 decimal
digits of precision.}
\label{f3}      
\end{figure}

Different network types exhibit significant differences in the
behaviour of damage spreading. This appears especially near
boundary of chaos and order, and is more intensive for $s=2$. 
In fig.~\ref{f3} we compare average damage size for initiation. It also contains 
the phenomenon of real fade-out in the first few steps. The shown points 
have 3 decimal digit of precision. All cases exhibit different behaviour
of damage spreading, despite the fact that 4,2 and 2,3 have the
same $w=1.5$, therefore we cannot limit ourselves to only one of
parameters $K$, $s$ or $w$. In addition to the different levels of damage 
equilibrium for different $s$ (which were shown earlier), now we encounter 
very different fade-out pattern in the first steps connected with network 
type and $s$.

\section{Complexity Threshold}

Let us investigate the evolution of distribution of damage size $d$ at
fadeout in dependency on system $N$ in more detail. Above we have
considered autonomous networks, but when we investigate a real
system, we can only observe its outside properties or a few points
inside, which we can describe as system external outputs.
It is similar to Ashby's  'essential variable' \cite{Ashby,ooKauf}.  
Therefore we suggest  simultaneous considering of effects of damage
on external network outputs. Let the number of output signals be fixed
as $m=64$. As a parameter analogous to damage size $d\in <0,1>$  we will 
use the Hamming distance - number of changed output signals $L\in <0,m>$
but without normalization. Distributions $P(d)$ and $P(L)$ should
be similar. Asymptotic value of $d$ which we named $d_{mx}$ and
asymptotic value $L_{mx}$ are simply dependent: $d_{mx}=L_{mx}/m$
but such a dependency is not valid during system growth and $L$ is
smaller than we would expect.

Evolution of these distributions starts for small $N$ as one peak
distribution, similar to distribution for ordered system $s=2$ and
$K=2$ when damage quickly fades out. Next, when $N$ is greater,
the second peak appears on the right long tail. It shifts to the
right and stops in some position of damage equilibrium. For
networks with feedbacks, before this right peak stops, between
peaks there appears a large period of practically exactly zero
frequency. Distribution $P(L)$ for different network types is shown in
fig 4.1.

\begin{figure}[ph]
\centerline{\psfig{file=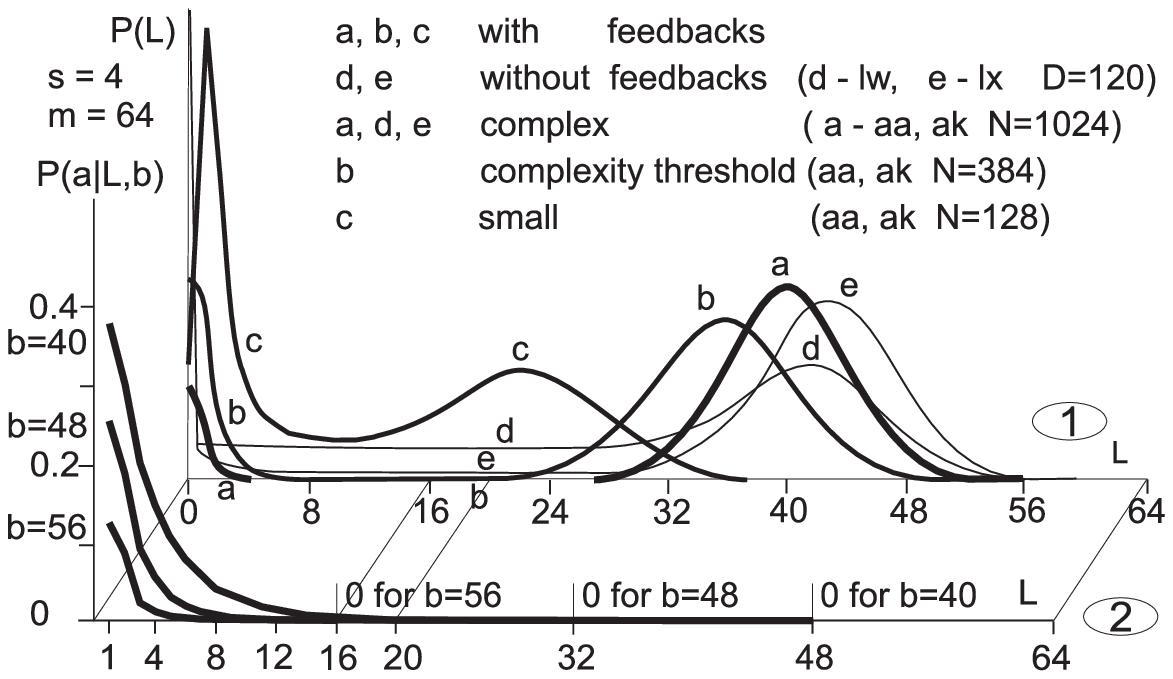,width=9cm}}
\vspace*{8pt}
%\begin{figure}
%\centering
%\includegraphics[height=5cm]{fig4.eps}
\caption{ Threshold of complexity (1) and small change tendency (2)
for $s=4$. When network is complex the main right peak of $P(L)$ is out of
range of acceptable cases for higher fitness $b$. (1). $P(L)$ distribution
of change size on the system outputs from simulations as the effect of 
random changes in the networks.
The base curve a - for complex network with feedbacks. It has two
peaks and exactly zero in-between. The same network on complexity
threshold - curve b, and below it, when network was still small  -
curve c. Size of the left peak strongly depends on network type,
here $aa$ network with small left peak.  
For networks without feedbacks $lw$ (most ordered) and $lx$ - curves d and e, 
the space between peaks has a small but visibly  non-zero value. 
(2). Calculated distribution of probability $P(a|L,b)$ of acceptance
by adaptive condition test, for three higher values of fitness $b$. 
For interesting higher $b$ the  $P(a|L,b)$
is significantly different from zero only for very small change
size ($L<20$ for $b=40$). To the right of the indicated points
$P(a|L,b)$ becomes exactly zero ($L=48$ for $b=40$).}
\label{f4}      
\end{figure}

\begin{figure}[ph]
\centerline{\psfig{file=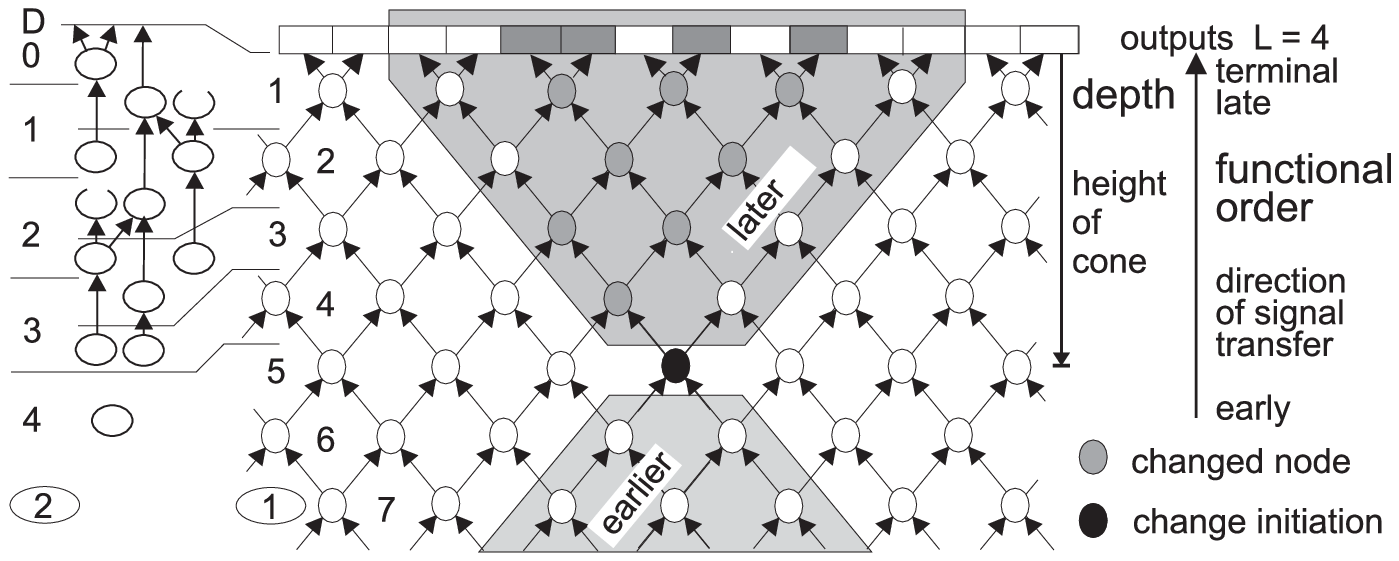,width=10cm}}
\vspace*{8pt}
%\begin{figure}
%\centering
%\includegraphics[height=4cm]{fig5.eps}
\caption{1.Cone of influence in an extremely simple network $lw$
without feedbacks. Black node divides the node set into three parts:
earlier and later nodes, which form the cone, and the rest - independent 
nodes. If the black node changes its state, then it becomes a source of damage,
which flows up to the outputs(grey nodes). If $w>1$ usually it is also
a cone (of damage). The depth of change initiation is the height of the
cone and suggests the average change size $L$. It is connected
with functional order and place description as early, late and
terminal as shown. 2.Depth definition used in $aa$ simulation and
its comparison to the shortest way to outputs used for $sf, ss$ and
levels number used for $lw, lx$. It is a structural and sequential
approximation of the functional order.}
\label{f5}      
\end{figure}

To understand the mechanism of this evolution let us consider an
extremely simple example of network `$lw$' of nodes with $K=k=2$ and functions
as in `$aa$', without feedbacks and with clear levels
of nodes, wrapped around a cylinder to remove the left and right ends. 
On each level there are 32  nodes connected according to the ordered pattern
like in fig.~\ref{f5}.1. For such an extremely simple network we can draw a cone of 
influence (fig.~\ref{f5}.1) which splits the network into three parts of nodes: nodes 
later than the selected one (which
are dependent on the selected node); earlier ones (which have an influence on
the selected one); and independent ones. To define sequence (earlier -
later) the sequence of signal flow and transformation depicted by
arrows of directed network is used. This is the functional order.

The cone of influence shows which nodes and outputs can be reached
by damage from a given source of damage, but not all of them will actually 
be reached by every particular case of damage. The affected ones create a 
smaller cone inside the later area.
If the later part of cone of influence does not include a part
of outputs, then the signals on these outputs cannot change as a result of 
damage.
The number of outputs inside the `later cone' depends on the cone height -
here it is measured in levels from outputs on the top down to the
source of damage, therefore we named it `depth' and denoted it by $D$.

If $D$ increases by one, then for $lw$ the number of later outputs
increases only by two. Let us decrease  order in this network and
define `$lx$' network where connections between neighbouring levels
are random. Now the number of later outputs increases two times
for the first few levels. For $s=4$, for $lw$ the right peak in
$P(L|D)$ achieves a stable position for damage initiation on depth
$D=78$, but for $lx$ it occurs much faster - at $D=26$. If we take
$s=8$, then the order also decreases and we need accordingly less
levels for peak stabilization for $lw$ $D=46$ and for $lx$ $D=13$.
For smaller $D$ the right peak in $P(L|D)$ shifts from the left to
$L_{mx}$ on the right with different tempo for both networks. To
obtain total $P(L)$ we must summarize $P(L|D)$ distribution for
all levels. In effect we obtain non-zero frequency in the whole
period from zero on the left to $L_{mx}$. For systems with more
levels the right peak in $P(L)$ grows and the contribution of constant
numbers of cases in-between peaks decreases.

For networks with feedbacks the notions of cone of influence, functional
order and depth dramatically (but not completely) lose their precision. We
can define depth as the smallest distance to outputs, but in
a different aspect this depth can be infinite if we consider the loops.
For an $lw$ network the maximum of the right peak in $P(L)$ achieves
80\% of $L_{mx}$ ($s=4$) for $N= 1696$, but for $lx$ network for
$N=320$ and for $aa$ network for $N$ equal to only $310$ (see table 1).

The stabilization (i.e. stop of shifting to the right during
network growth) of the right peak in $P(L)$ and in $P(d)$ and its
parameters $N$, $s$ and structure in the shown examples, correctly 
correspond to our intuitive notion of complexity. (If
similar states of the system create very different effects, we
must know much more to predict these effects.) We define them as
complexity threshold, however for networks with feedbacks it is also the 
chaos (in Kauffman's meaning) threshold. In networks containing feedbacks 
it is accomplished by appearance of
practically zero frequency in-between peaks but these two
phenomena do not appear exactly simultaneously. They can be used as
other variants of complexity threshold.

\begin{table}[ht]
\tbl{Complexity threshold for different parameters and criteria}
{%\centering \caption{Complexity threshold for different parameters
%and criteria}
\label{ta1}     
\begin{tabular}{@{ }r@{ \ \ }r@{ \ \ }r@{ \ \ }r@{ \ \ }r@{ \ \ }r@{ \ }r@{ \ }r@{ \ }r@{ \ }r@{ \ }r@{ \ \ }r@{}}
\hline\noalign{\smallskip}
network & aa4,2 & ak4,2 & er4,2 & ss4,2 & sf4,2 &  ak16,2 & ak64,2 & ak4,3 & sf16,2 & sf64,2 & sf4,3 \\
\noalign{\smallskip}\hline\noalign{\smallskip}
0oc. N & 384 & 384 & 768 & 1536 & 2048 & 128 & 96 & 96 & 512 & 320 & 768 \\
0oc. \%$d_{mx}$ & 92 & 93 & 91 & 90 & 85 & 91 & 90 & 71 & 85 & 82 & 84 \\
0oc. \%$L_{mx}$ & 85 & 85 & 85 & 74 & 73 & 74 & 67 & 79 & 63 & 55 & 58 \\
N 90\%$d_{mx}$ & 340 & 384 & 730 & 1810 & 4600 & 128 &  96 &  56 &  900 &  512 & 1720 \\
N 80\%$L_{mx}$ & 310 & 280 & 620 & 2000 & 3850 & 200 & 190 & 110 & 1390 & 1160 & 5000 \\
\noalign{\smallskip}\hline
\end{tabular}}
\end{table}

We have found parameters for appearance of all three phenomena
(table ~\ref{ta1}) as three criteria of complexity threshold. First three
rows show zero occurrence (0oc.). Percent of  $d_{mx}$ and
$L_{mx}$ is shown for comparizon of zero occurrence criterion to two
others shown  below. Networks $er$, and $ss$ always lay between
$ak$ and $sf$, therefore we simulate them only for $s=4$ and $K=2$
where in-between peaks there appears an area of near-zero frequency 
when position of right
maximum in $P(d)$ is on 90\% of $d_{mx}$ and in $P(L$) on 80\% of
$L_{mx}$. If $s$ or $K$ grows then this coincidence no longer occurs.
The scale-free network is an extreme case and it achieves all of
these criteria much slower than $ak$. Networks $aa$ and $ak$
differ in $P(d)$ distribution value but in $P(L)$ they are very
similar or even identical. Much more data is used for these
conclusions, which we cannot include here due to limited space,
they will be described separately. %                          \cite{C2}. 
We do not show error ranges either, which are circa 3\%
for most of stability positions and ca 20\% for $N$ of zero
appearance but they are not important due to large difference of
values. For comparison: `How should complexity scale with system size?' considered in a theoretical way by Olbrich et al. \cite{Olbrich}.   

\section{Tendency as Difference in Distribution in Adaptive and Free
Processes, Small Change Tendency}

To investigate adaptive evolution we must define fitness $b$,
which should not decrease. Fitness can be defined using states of
all nodes (this way was applied by Kauffman), or using effects of system function which are accessible outside of the system, i.e. external output signals (as 'essential variable'). 
We prefer the second solution. The simplest method is
to arbitrarily define some ideal vector of  $m$ signals and to
compare it to the vector of system output signals. Let fitness $b$ be
the number of identical signals in both of these vectors. This is
a common method.

System is changed, e.g. by addition or removing of a node (fig.~\ref{f2}).
After such a change it has a certain fitness which describes its state.
Consecutive system states of fitness counted by $t$ are delimited by
changes of system construction. This creates a process. All changes creating an adaptive
process meet the adaptive condition $a \equiv (b_{t+1} \geq b_t)$
which is used to eliminate (not accept) some random changes in the
simulation. We will compare the adaptive process to the free
process, which accepts all random changes, as they are drawn.

The tendency is a difference between probability distribution $P(X|a)$
of change parameter $X$ for an adaptive process and $P(X)$ for a
free process. Note, $X$ must describe a change, it cannot be
a state parameter.

>From Bayes:  $P(a) P(X|a) = P(a|X) P(X)$

$P(a)$ is a constant; therefore, a tendency is shown by $P(a|X)$ .
As we can see, we do not have to know $P(X)$ to know the tendency.
It is enough that for different $X$, $P(a|X)$ is different.
However, in the structure development of complex system $P(X)$ is
important due to other causes.

It is useful to introduce a general parameter of process advancement
connected to the $t$ count, let it be denoted  by $g$. It is the
state describing parameter, e.g. it can be $t$ or $b$. Similarly
we can find, that tendency is described by $P(a|X, g)$ and we will
use this form later.

The first, very simple but very important one, is the tendency to collect in
adaptive process much smaller changes than the changes creating a free
process. We named it `small change tendency'. It is the base of
mechanisms of  all the structural tendencies investigated later, which are more interesting. 
This tendency is a different view on underlying by Kauffman `structural stability' as a condition of adaptive evolution \cite{ooKauf}.

If we limit our consideration only to output vectors and 
assume, that each signal changes independently, then we can
calculate $P(a|L,b)$ for given $s$. This is the form obtained above, which 
indicates a tendency when $P(a|L,b)$ really depends on $L$. For higher $b$
only very small changes are acceptable - see fig.~\ref{f4}.2, where it is
compared to $P(L)$: for networks above the complexity threshold all
cases from the right peak cannot be accepted by the adaptive condition test.

\section{Structural Tendencies of Terminal Modifications and Terminal
Predominance of Additions}

If we consider both: small change tendency and dependency of change
size $L$ on depth $D$ in the construction of cone of influence then
we can expect `terminal modifications' tendency known in classic
developmental biology \cite{Naef}. Depth $D$ is a structural
approximation of functional order which creates the cone of influence and
time of ontogeny stages. However, the cone of influence is well
defined in a system without feedbacks, but if feedbacks are present,
then it can only be a premise. The answer can only be obtained by
simulation, where we should check $P(a|D)$. For $aa$
investigations we have used a special, more adequate definition of
depth $D$ (fig.~\ref{f5}.2), but it is not to be  applied to various $k$
in other networks.

To investigate the system development, this system must have
the ability to grow, therefore in the set of possible changes there should be
addition of a new node. For higher adequateness  there should also be
the removing of a node. Both of them should be drawn randomly, but
the sets of possibilities for such a draw cannot be the same. Removing
can only be drawn from nodes present in the network, but addition
has a much larger set of possibilities. This difference can create some
difference of acceptance probability for additions and removals in different areas of
network, which differ with respect to modification speed in effect of 
the terminal modification tendency. 

Note, that additions and removing transform a particular system into another 
one. It is a walk in the system parameters space in the Kauffman approach 
\cite{ooKauf} but in our approach we can see important differences between 
probability to adaptive move using addition and probability using removal, %p11
and we can distinguish between various areas in the system body. However, 
similarly to Kauffman, we have a close similarity between the effects of small changes which change the system to another (additions, removing of nodes) and the ones which affect the system state only (changes of state of a node). In our case it is not an assumption but a simulation result.

\begin{figure}[ph]
\centerline{\psfig{file=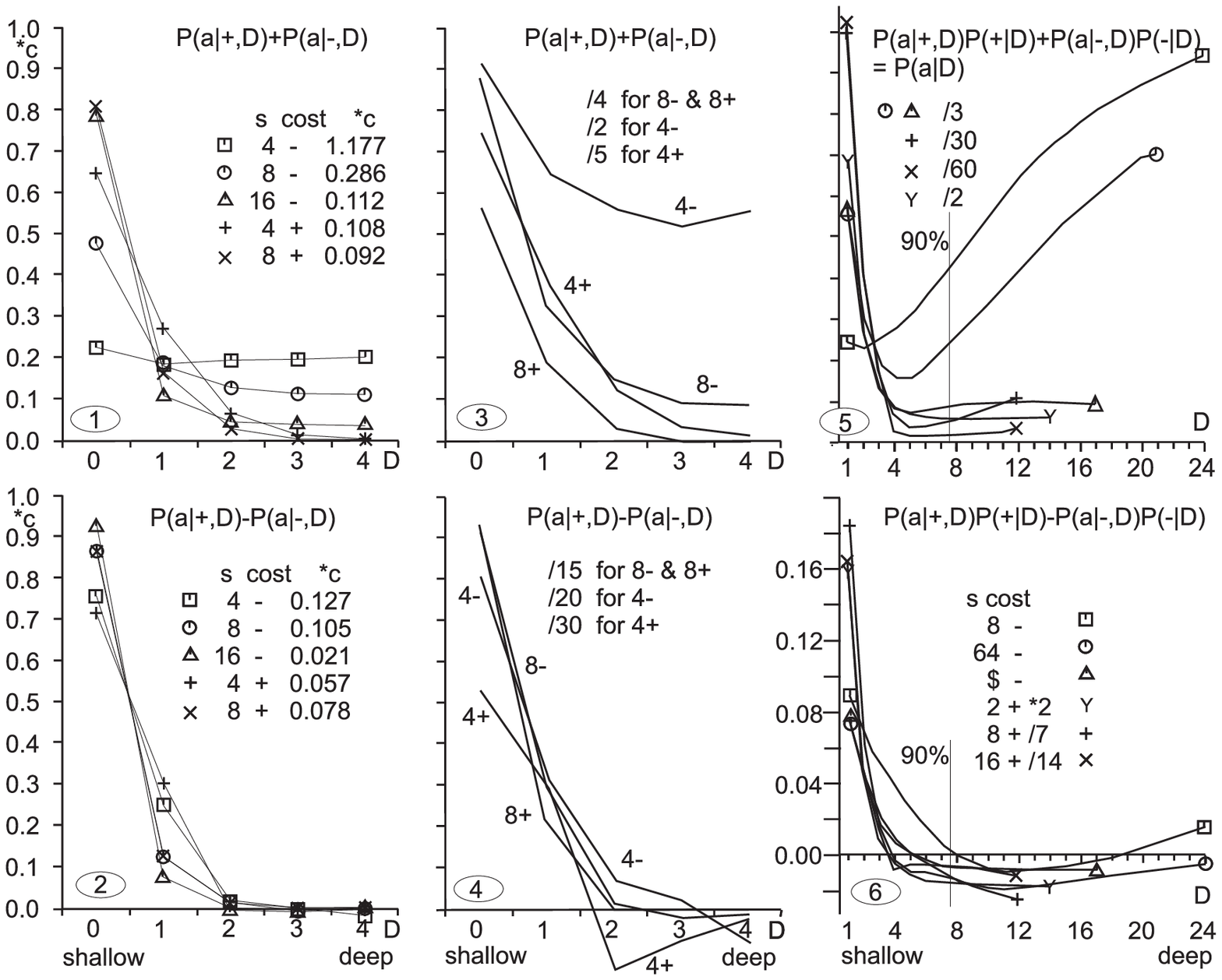,width=12cm}}
\vspace*{8pt}
%\begin{figure}
%\centering
%\includegraphics[height=9.4cm]{fig6.eps}
\caption{The main result of simulations, (scaled by c for
comparison). On the left (1,2) for $aa$, in the middle (3,4) for
aggregate without feedbacks `$an$' and on the right (5,6) for
Kauffman networks of variable node degree $k$ - scale-free $se$
and single-scale $ss$. The results for $ss$ and $se$ are so
similar, that a single curve is used for both of them. Dependency on
depth $D$ as a structural measure of functional order. Definition of
$D$ for 1-4 in fig.~\ref{f5}.2, for 5 and 6 it is the shortest way to
outputs. On the top (1,3,5): Terminal modifications and
conservation of early area tendency. At the bottom (2,4,6):
Balance of addition and removing: terminal predominance of
additions (over removing) tendency and tendency of simplification
(predominance of removing over addition) of the early parts.}
\label{f6}      
\end{figure}

\begin{figure}[ph]
\centerline{\psfig{file=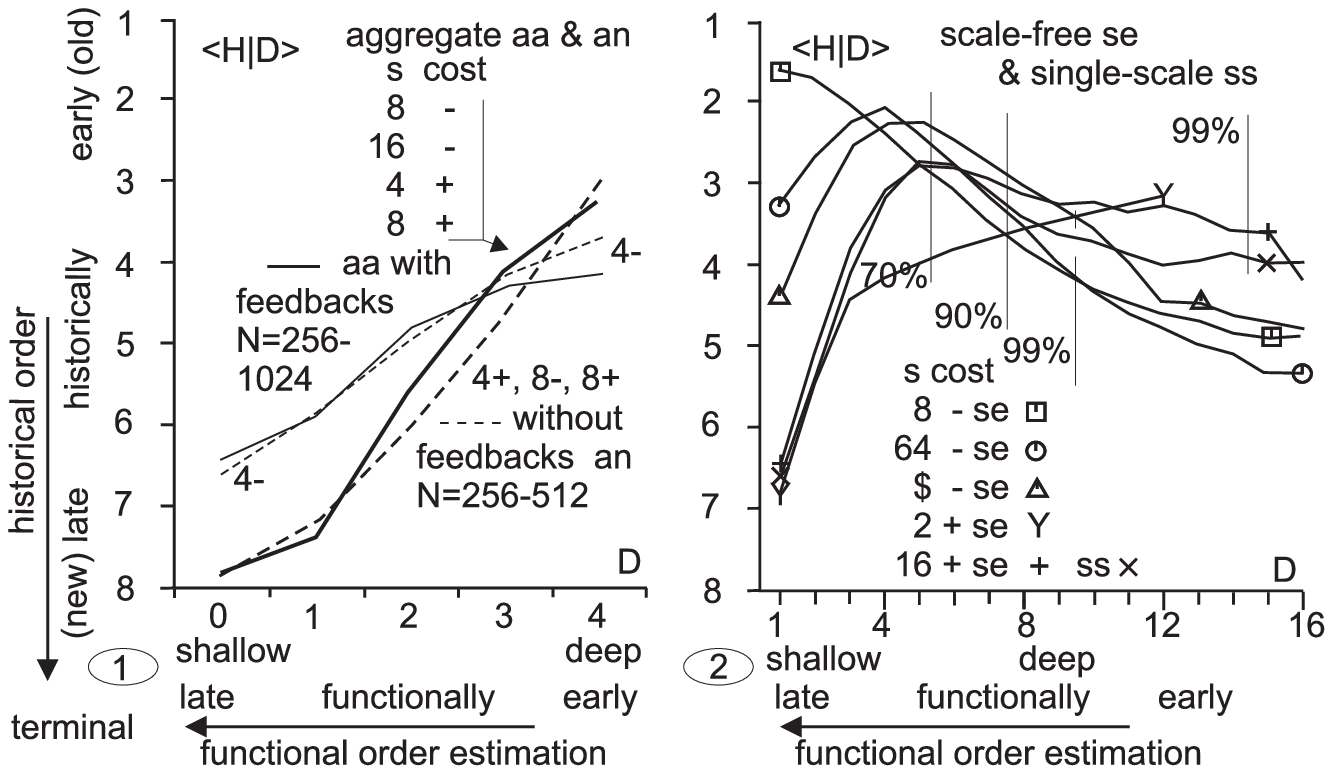,width=10cm}}
\vspace*{8pt}
%\begin{figure}
%\centering
%\includegraphics[height=5.6cm]{fig7.eps}
\caption{Similarity of historical $H$ and functional $D$ order.
$H$ is a sequence of addition of given node to the network. The
same set of simulations as in fig. ~\ref{f6}. On the left (1) for
aggregate with feedbacks ($aa$) - continuous lines, and without
feedbacks ($an$) - dashed lines. On the right (2) for Kauffman
networks with feedbacks ($se$ and $ss$), containing $k=0$ and
$k=1$. Here $D$ is the shortest distance to outputs and can be
large, but with small probabilities, therefore the boundaries of
these probabilities are shown.}
\label{f7}       
\end{figure}

One of the typical cases of removing is a removing of a `transparent' node which does not change signals of the remaining nodes. It especially occurs when a `transparent' node is just added.
Such a case may have different interpretations. Some of them suggest
forbidding transparent addition. We introduce such forbidding
using strict inequality in the adaptive condition for additions,
and weak inequality for removals. This is equivalent to a cost
function for additions of a new automaton. 
This simple `cost' condition appears very strong which is easy to understand
 - newly connected nodes must lie closely near assessed outputs to influence 
at least one signal of system outputs but not much more. 
It eliminates additions on longer distance in both regimes - in chaotic one because change of outputs is too large and in ordered one because damage fadeout without affecting outputs (fig. ~\ref{f6} and ~\ref{f7}).

Similarly to above, we have checked these assumptions in simulation
for different network types. Random $er$ network cannot grow
and therefore we do not use it for these experiments. 
For important $sf$ network a problem appears for removing, which creates $k=0$
nodes. Such a node cannot come back into play and creates
a dummy network where most of nodes have $k=0$. To correct this
situation we add to link drawing also node drawing like in $ss$
network and we obtain connection proportionality to $k+1$ for
modified $sf$ type which we name $se$. It is a little similar
to Ref.~\refcite{Dorogovtsev2000}.  However this modification, especially
important for case without cost, modifies $P(k)$ which becomes more like
for $ss$.

For correct description of phenomena similar to transparent
addition and removing we now use node functions in our algorithm. 
As it can be expected, $se$ and $ss$ networks need much higher $s$ to
obtain typical phenomena for $aa$ and $ak$ networks. 
To create damage spreading inside the network as for very high $s \sim 1800$ we
use a special function and $s=64$, denoted in figures as \$.

The main results are shown in fig.~\ref{f6}.  They are obtained in three
series of simulations, all with fixed $K=2$. First for $aa$, where
clear and strong tendencies of terminal modifications and
conservation of deeper part (fig.~\ref{f6}.1), and terminal predominance
of addition over removing and simplification of deeper parts are
obtained (fig.~\ref{f6}.2). Only the case of $s=4$ without cost appears to
be extreme (still too small $s$). Next we have investigated the
question: how important are feedbacks in the mechanisms of these
tendencies? For network `$an$' (similar to $aa$ only devoid of
feedbacks), we have obtained similar effects (fig.~\ref{f6}.3 and 6.4). In
the last series we ask: are there the same tendencies in the
contemporarily preferred networks $se$ and $ss$, with various node
degree? The answer is shown in fig.~\ref{f6}.5 and 6.6 - yes, they are there,
but in these networks there are much more interesting phenomena
connected with $k<2$ and hubs. One of them is deep fadeout
tendency - in deeper parts of such networks nodes of $k<2$ are
collected, hubs take their place at a small depth, but not very
small. If cost is absent, then this tendency is very
strong for $s<64$ and blocks other tendencies. In all cases, probability of
addition and removing was the same (before elimination) and networks grow, but for $ss$
and $se$ if cost is present, then the networks do not grow. 
This is an effect of deep fadeout for removing. In order to grow
they need a very small percentage of removing in the tested changes, ca. 1\%.
During system growth the number $m=64$ of outputs is constant. The volume %p13
of small depth, where damage easier reach outputs, quickly is fulfilled and 
any addition of new nodes (wherever it happens) only expands deeper the part of the 
system where the way to output is long and the damage has higher chance to fade 
out without reaching the outputs.  Therefore additions are there not 
acceptable (cost) but removals are accepted with higher probability and 
growth of the network stops on the certain level of $N$.

High terminal predominance of addition over removing corresponds
with Weismann's `terminal additions' regularity
\cite{Weismann1904} and in the same way creates similarity of
historical and functional order (fig.~\ref{f7}). The historical order is a
sequence of connections of given node to the network. It is the
main element of the famous `recapitulation of the phylogeny in the
ontogeny' regularity known also as the Haeckel's `biogenetic law'
\cite{Haeckel}. The stability of function is a completion of this similarity of orders. 
We also observe it in our simulations.
It is a pity that recapitulation has died in 1977
\cite{Wilkins2002,Gould1977} because of the lack of explanation,
before our proving of its mechanisms, despite our attempts
\cite{paris}. It is obvious, that this short announcement is not
enough to  reanimate it, therefore much longer descriptions are
under preparation.  %                                       \cite{C1,C2,C3}.

\section{Conclusion}

The way to a mechanism of recapitulation is long and contains (in
opposite direction): tendency of terminal predominance of addition,
which needs tendency of terminal modifications. We have observed
them in simulation in a few different network types as effects
of adaptive condition of network growth, over a certain complexity
threshold. This threshold is observed the in damage size distribution
on network external outputs. To model damage spreading we use
functioning, directed networks, e.g. Kauffman networks, but we use
more than two equally probable signal variants, therefore they are
no longer the Boolean networks. Such an assumption (argued using
interpretation) places the considered systems in the chaotic area, far
from `order' and `phase transition to order', in
addition the complexity threshold guards this assumption. 
In comparison to the Kauffman model we introduce two new elements. Firstly, we allow  $s>2$ (more than two equally probable signal variants), which keeps our system in the chaotic area (unlike `internal homogeneity' $P$). Secondly, we introduce external outputs which we use for differentiation of system body 
areas and as `essential variables' for fitness definition.  %p13
We find  `structural stability', important for adaptive evolution, in our `small change tendency' which causes differences in elimination between various areas of network body. These differences lead to `structural tendencies'. The famous Kauffman conclusion from his model `We shall find grounds for  thinking that the ordered regime near the transition to chaos is favored by, attained by, and sustained by natural selection' seems not applicable to our model.
Because our definition of fitness uses output signals, we omit the problem 
of local optima, which are absent here and the problem of `complexity 
catastrophe' expected by Kauffman.
Simplifications in `fitness landscape' and especially in algorithm in the aspect of attractors allow us to investigate a large, new and interesting area of `structural tendencies' which is waiting for a more mathematical description.

%%%%%%%%%%%%%%%%%%%%%%%%%%%%%%%%%%%%%%%%%%%%%%%%

%\printindex
\end{document}